\documentclass[%
preprint,
showpacs,
amsmath,amssymb,
aps,prl,
]{revtex4-1}

\usepackage{graphicx}
\usepackage{dcolumn}
\usepackage{bm}
\usepackage{subfigure}

\begin{document}
\hyphenation{eq-ua-tions diff-er-ent only sce-nario also however equi-librium fila-ment results works re-mains two still account University} 

\title {Roughness-facilitated local 1/2 scaling does not imply the onset of the ultimate regime of thermal convection}



\author{Xiaojue Zhu$^{1}$}
\email{xiaojue.zhu@utwente.nl}
\author{Richard J. A. M. Stevens$^{1}$}

\author{Roberto Verzicco$^{2,1}$}

\author{Detlef Lohse$^{1,3}$}
\email{d.lohse@utwente.nl}

\affiliation{$^1$Physics of Fluids Group, MESA+ Institute and J. M. Burgers Centre for Fluid Dynamics, University of Twente, P.O. Box 217, 7500AE Enschede, The Netherlands\\
$^2$Dipartimento di Ingegneria Industriale, University of Rome `Tor Vergata',
Via del Politecnico 1, Roma 00133, Italy\\
$^3$Max Planck Institute for Dynamics and Self-Organization, 37077 G\"ottingen, Germany}




\begin{abstract}
In thermal convection, roughness is often used as a means to enhance heat transport, expressed in Nusselt number. Yet there is no consensus on whether the Nusselt vs. Rayleigh number scaling exponent ($\mathrm{Nu} \sim \mathrm{Ra}^\beta$) increases or remains unchanged. Here we numerically investigate turbulent Rayleigh-B\'enard convection over rough plates in two dimensions, up to $\mathrm{Ra}=10^{12}$. Varying the height and wavelength of the roughness elements with over 200 combinations, we reveal the existence of two universal regimes. In the first regime, the local effective scaling exponent can reach up to 1/2. However, this cannot be explained as the attainment of the so-called ultimate regime as suggested in previous studies, because a further increase in $\mathrm{Ra}$ leads to the second regime, in which the scaling saturates back to a value close to the smooth case. Counterintuitively, the transition from the first to the second regime corresponds to the competition between bulk and boundary layer flow: from the bulk-dominated regime back to the classical boundary-layer-controlled regime. Our study clearly demonstrates that the local $1/2$ scaling does not signal the onset of asymptotic ultimate thermal convection.

\end{abstract}


\maketitle


Thermal convection plays an important role in a wide range of natural and industrial environments and settings, namely from astrophysical and geophysical flows to process engineering. The paradigmatic representation of thermal convection, Rayleigh-B\'enard (RB) flow, in which a fluid is heated from below and cooled from above, has received extensive attention over the past decades \cite{ahlers2009,lohse2010,chilla2012}.
One of the major challenges in the studies of RB convection is to determine the scaling relation of the Nusselt number ($\textrm{Nu}$), which is the dimensionless heat flux, with the Rayleigh number ($\textrm{Ra}$), which is the dimensionless temperature difference between the two plates, expressed as $\textrm{Nu} \sim \textrm{Ra} ^{\beta}$. 

Assuming that the heat transport is independent of the cell height and governed by the viscous boundary layers (BL), Malkus \cite{malkus1954} derived that $\beta=1/3$. Later, Grossmann \& Lohse \cite{grossmann2000,grossmann2001} showed that there is no pure scaling but smooth transitions from BL to bulk dominated regimes. However, for large $\textrm{Ra}$ when the whole system is highly turbulent and laminar type BLs do not play a role anymore, Kraichnan \cite{kraichnan1962} postulated that the flow reaches the ultimate state in which $\textrm{Nu}$ scales according to $\textrm{Nu} \sim \textrm{Ra}^{1/2} \mathrm{ln} \textrm{Ra}^ {-2/3}$, with $\mathrm{ln} \textrm{Ra}^ {-2/3}$ as the logarithmic correction term. This ultimate regime was also predicted by Grossmann \& Lohse \cite{grossmann2011}, who modelled this log-behavior with an effective scaling exponent of $\beta \approx 0.38$, for $\textrm{Ra}$ around $10^{14}$. Indeed, for $\mathrm{Ra} \approx 10^{14}$ the onset of such a regime with strongly enhanced heat transport has experimentally been confirmed \cite{he2012a,he2012b}. The logarithmic correction term becomes irrelevant for very large $\textrm{Ra}$. In this asymptotic ultimate regime $\beta=1/2$, and the heat transport is independent of viscosity and therefore the scaling can be extrapolated to arbitrarily large $\textrm{Ra}$, as those present in both geophysical and astrophysical flows. This asymptotic ultimate $1/2$ scaling has numerically \cite {lohse2003,calzavarini2005,schmidt2011} and experimentally \cite{gibert2006,cholemari2009,pawar2016} been observed in the so-called `homogeneous' or `cavity' RB turbulence, where no BLs are present.

Clearly, it is the interplay between BL and bulk flow which determines the effective scaling exponent \cite{grossmann2000}. To better understand the role of the BLs, it is important to alter the boundaries to probe how the system responds. Hence, much attention has been paid to RB turbulence over rough surfaces. Another motivation obviously is the fact that the underlying surfaces of real-world applications of thermal convection are always rough. Though it is generally agreed that roughness enhances the absolute value of $\textrm{Nu}$, there is no consensus on whether the {\it scaling exponent} increases with roughness \cite{ciliberto1999,stringano2006,roche2001,qiu2005,tisserand2011,wei2014,salort2014,sebastian2014,toppaladoddi2017,xie2017} or remains unchanged \cite{shen1996,du2000,shishkina2011} as compared to the smooth counterpart. For example,  Shen {\it et al.}  \cite{shen1996} found that $\textrm{Nu}$ increased by 20\%, whereas the exponent $\beta$ did not change upon using rough surfaces made of regularly spaced pyramids. Roche {\it et al.} \cite{roche2001} obtained an increase of $\beta$ to approximately 0.51 by implementing V-shaped
axis-symmetrical grooves both on the sidewalls and horizontal plates, which was interpreted as triggering ultimate regime scaling 1/2. Very recently, a roughness induced effective 1/2 scaling was found in the range of $\textrm{Ra}=[10^8, 10^9]$. This was again explained as the attainment of the ultimate regime \cite{toppaladoddi2017}. However, it is surprising that the ultimate regime can be found at such low $\textrm{Ra}$ since theories show that the ultimate regime 1/2 scaling can only be observed asymptotically when $\textrm{Ra}$ reaches infinity \cite{grossmann2011,goluskin2016}. In addition, the 1/2 scaling observed might possibly be due to a crossover between rough surfaces from a regime with a groove depth less than the BL thickness to a regime where the groove depth is larger than the BL thickness, as speculated in Ref. \cite{ahlers2009,tisserand2011,xie2017}.

In this study, we will unify these different views. For this, we perform direct numerical simulations (DNS) of turbulent RB convection over sinusoidally rough plates in two dimensions (2D), adopting the same roughness configuration as in Ref. \cite{toppaladoddi2017}. The effects of roughness on heat transport are presented by varying the heights $h$ and wavelengths $\lambda$ of the rough elements independently. Even though 2D RB differs from three dimensional (3D) RB in terms of integral quantities for finite $\textrm{Pr}$ \cite{schmalzl2004,poel2013}, the theoretical arguments for the scaling relations are similar \cite{grossmann2011,goluskin2016}. Indeed, 2D simulations are much less time consuming than 3D and can help us pushing forward to $\textrm{Ra}=10^{12}$ and $\textrm{Nu}\sim \mathcal{O}(10^3)$ with roughness. This key
extension to large $\textrm{Ra}$ unravels the physical origin of the 1/2 regimes observed in Ref. \cite{toppaladoddi2017}.

The simulations were performed using an energy conserving second-order finite-difference code \citep{verzicco1996,vandepoel2014} converted to 2D, in combination with an immersed-boundary method \citep{fadlun2000} to track the boundaries of the rough elements. The code has been extensively validated through prior investigations, with \cite{stringano2006} and without roughness \cite{vandepoel2014}. No-slip conditions were used for the velocity and constant temperature boundary conditions for rough bottom and top plates. Periodic boundary conditions were employed on the horizontal sidewalls. The control parameters of the RB system are the Rayleigh number $\textrm{Ra}=\alpha g \Delta (L-h)^3/(\nu \kappa)$ and the Prandtl number $\textrm{Pr}=\nu/\kappa$, where $\alpha$ is the thermal expansion coefficient, $g$ the gravitational acceleration, $\Delta$ the temperature difference between the top and bottom plates, $L$ the height of the fluid domain without roughness, $h$ the height of the roughness element, $\nu$ the kinematic viscosity and $\kappa$ the thermal diffusivity of the fluid, respectively. The reason to choose $L-h$ here for the rough cases as the characteristic length is that it resemble the height between the two smooth plates where the same volume of fluid occupies. The other flow quantities are nondimensionalized by the temperature difference $\Delta$ and the free fall velocity $U=\sqrt{\alpha g \Delta (L-h)}$. In all simulations, $\textrm{Pr}=1$ and the aspect ratio $\Gamma=D/L=2$, where $D$ is the width of the domain. With this $\Gamma$, the heat flux
approximates the heat flux at an infinite aspect ratio \cite{johnston2009}. Three roughness heights $h$ were chosen, $h/L=$0.05, 0.1 and 0.15. For each amplitude, the wavelength of roughness $\lambda/L$ was varied from 0.05 to 0.7. For each combination of wavelength and amplitude, we performed simulations in the range of $\textrm{Ra}=[10^8, 10^{12}]$. In total, 205 cases were simulated. Adequate resolution was ensured for
all cases, with at least 200 free fall time units. At $\textrm{Ra}=10^{12}$, $14336\times7168$ grid points were used.  
$\textrm{Nu}$ is calculated from $\textrm{Nu}=\sqrt{\mathrm{Ra}\mathrm{Pr}}\left <u_z \theta \right >_A -\left <\partial_z \theta \right >_A$, where $u_z$ denotes the instantaneous vertical velocity, $\theta$ the temperature, and $\left <\cdot \right>$ the average over any horizontal plane. 


 \begin{figure}

     \includegraphics[width=3.0in]{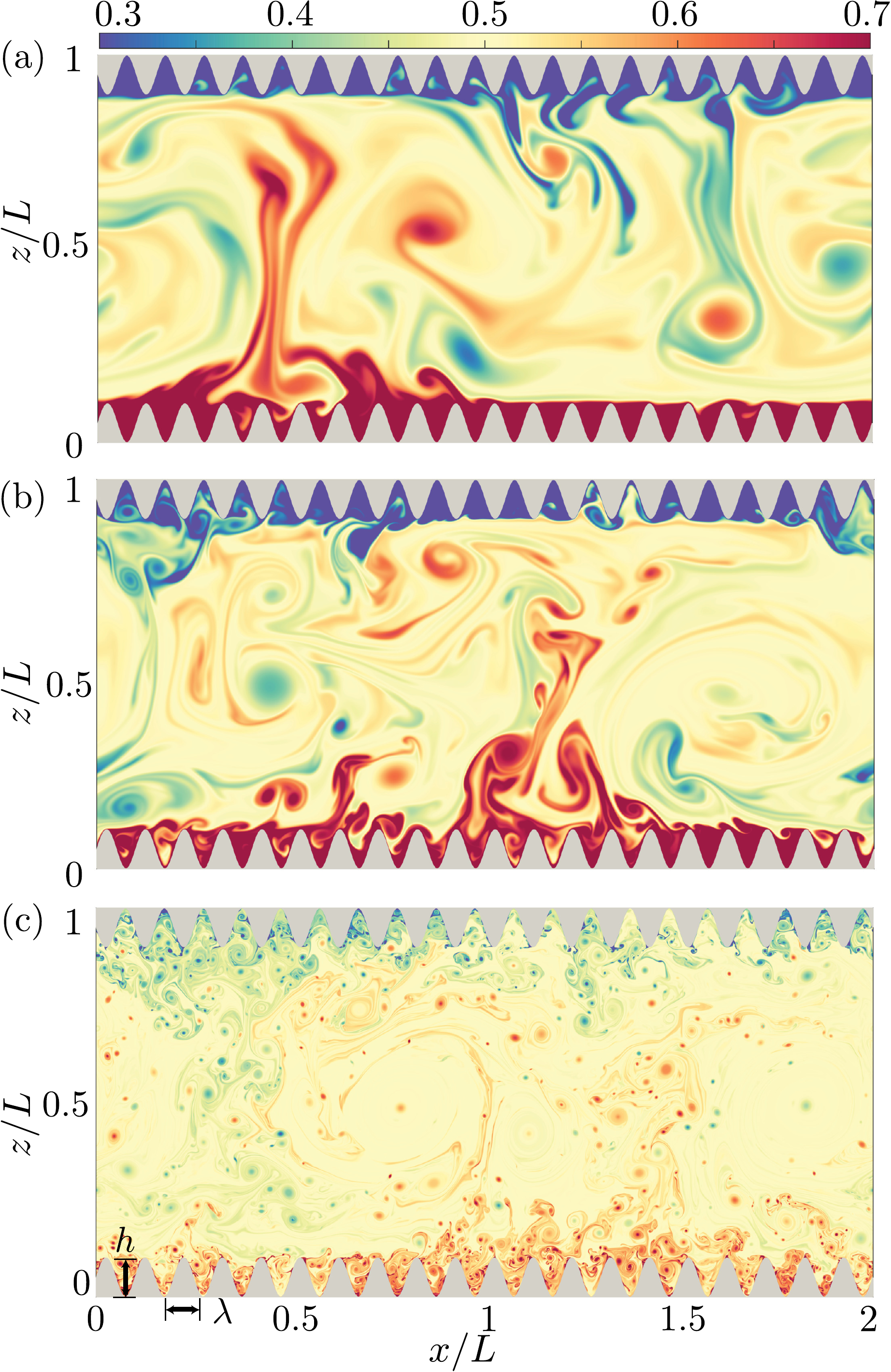}
    
      \caption{The instantaneous temperature fields for $\lambda/L=0.10$ and $h/L=0.10$ at (a) $\mathrm{Ra}=2.2\times10^8$, (b) $\mathrm{Ra}=2.2\times10^9$ and (c) $\mathrm{Ra}=7.3\times10^{11}$, where $\lambda$ is the wavelength and $h$ the height of the roughness. The three plots share the same colormap.}
\label{fig1}
\end{figure}

\begin{figure}[htbp]

     \includegraphics[width=3.in]{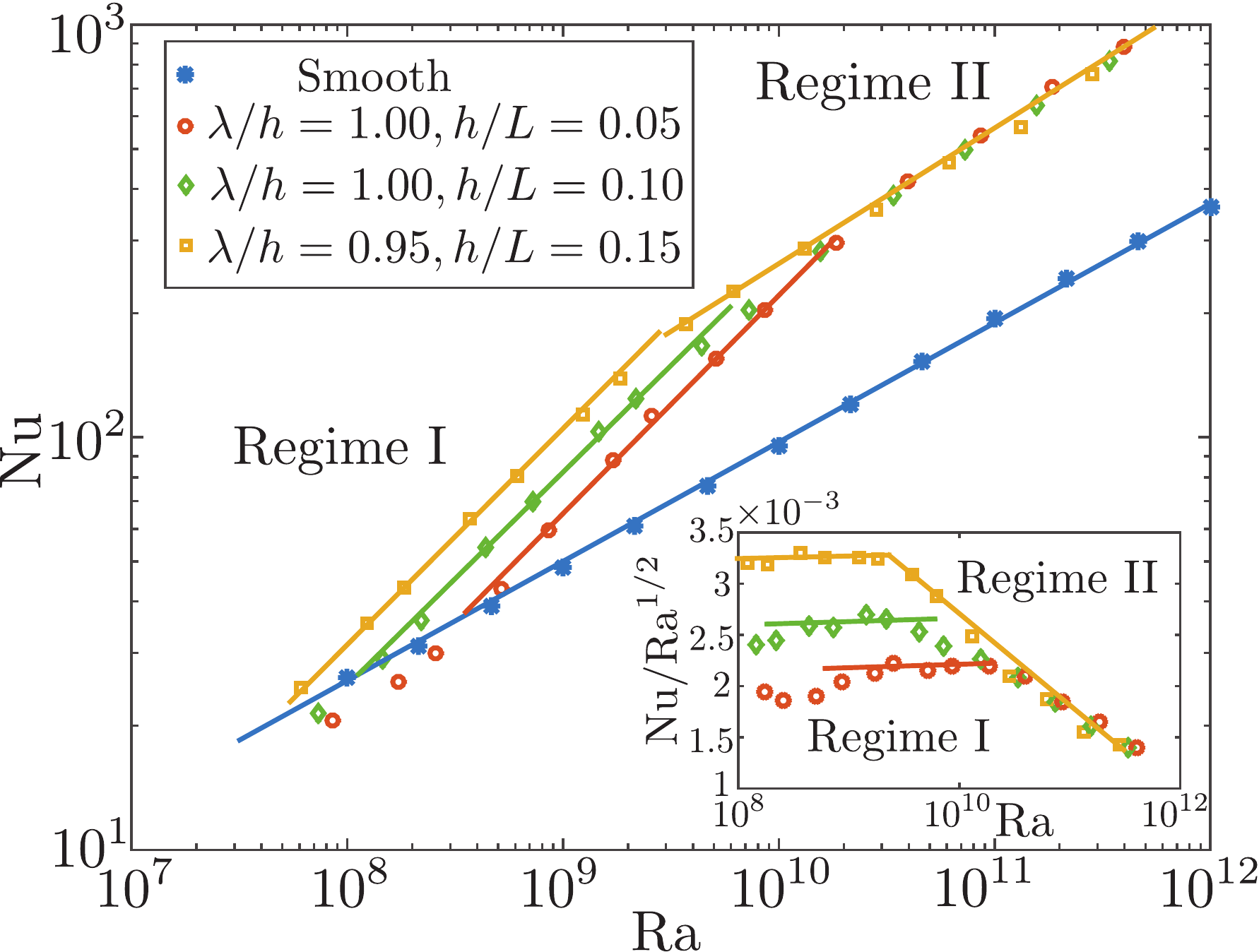}
    
      \caption{Nu(Ra) for rough cases of aspect ratio $\lambda/h \approx 1$ at $h/L=0.05, 0.10$ and $0.15$, in comparison to the smooth case, for which the scaling is exponent is $\beta=0.29\pm0.01$, crossing four decades. For the rough cases, although the heights of the rough elements are different, two regimes can be identified: Regime I, $\beta=0.50\pm0.02$ and Regime II, $\beta=0.33\pm0.01$. The inset shows the compensated plot and the plateau demonstrates the robustness of 1/2 scaling in Regime I.}
\label{fig2}
\end{figure}

\begin{figure*}

     \includegraphics[width=5.6in]{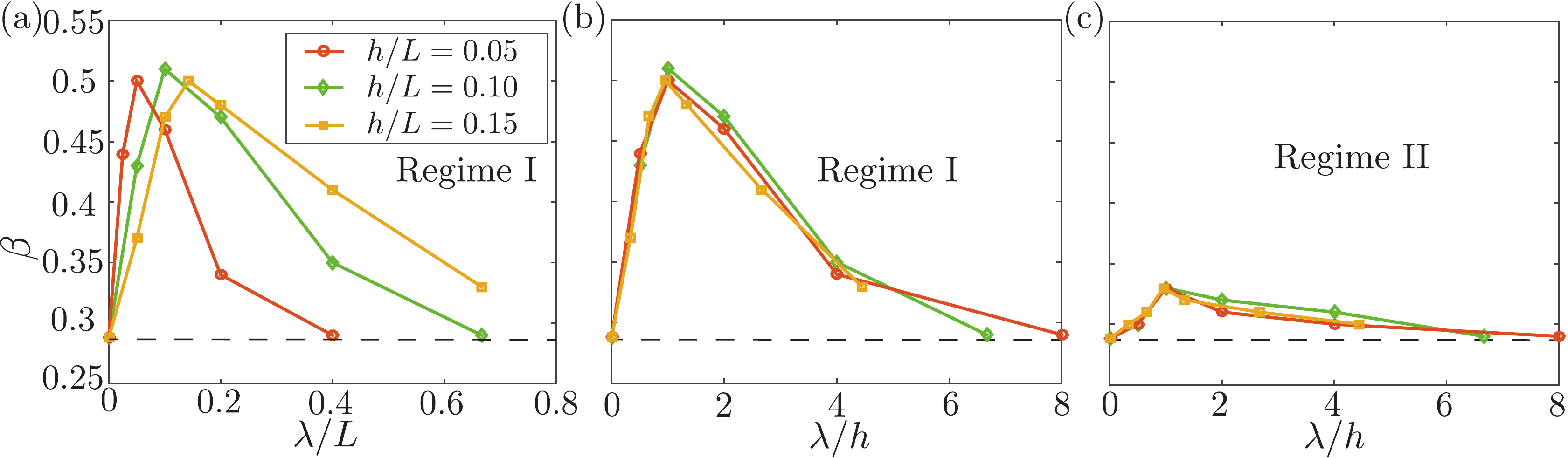}
    
      \caption{The $\mathrm{Nu}$ vs. $\mathrm{Ra}$ effective scaling exponents $\beta$ in Regime I and Regime II as a function of  (a) roughness wavelength $\lambda$ in Regime I, and (b)(c) aspect ratio $\lambda/h$ in Regime I and II, respectively. Note that both $\lambda=0$ and $\lambda=\infty$ correspond to the smooth plate case (dashed line).}
\label{fig3}
\end{figure*}

We begin by comparing the temperature field with increasing $\textrm{Ra}$ (see Fig. \ref{fig1}), for a fixed set of roughness parameters ($\lambda/L=0.1$ and $h/L=0.1$). Here we stress the plume morphology inside the cavity regions between the rough elements. For the two smaller $\textrm{Ra}=2.2\times10^8$ and $\textrm{Ra}=2.2\times10^9$, thermal plumes are mainly generated from the tips of the rough elements and are detached towards the directions of the large scale rolls, while in the cavity regions the flow is viscosity dominated. Note that below $\mathrm{Ra}=2.2\times10^8$, roughness elements are submerged inside the thermal boundary layer whereas $\textrm{Ra}=2.2\times10^9$ is already one order higher. In comparison, at $\textrm{Ra}=7.3\times10^{11}$, plumes are not only generated at the tips but also at the sloping surfaces of the rough elements. Inside the cavities, the detached plumes mix the fluid vigorously, making the flow there more turbulent. These observations suggest that even after the rough elements protrude beyond the thermal BL, the flow structure is essentially similar for one decade of $\textrm{Ra}$ while it changes drastically only when further increasing $\textrm{Ra}$ so that the flow inside the cavities becomes turbulent.

Having seen the effect of roughness on the flow structure, we now systematically explore the heat transport as a function of $\textrm{Ra}$, covering more than four decades. The resulting $\mathrm{Nu(Ra)}$ dependences with the same roughness aspect ratio $\lambda/h \approx 1$ for different roughness heights are displayed in Fig. \ref{fig2}. The smooth case follows an effective scaling exponent $\beta=0.29$, which is in very good agreement with previous studies \cite{poel2013,poel2014}. With the introduction of roughness, two universal regimes can be identified. When the roughness elements protrude the thermal BL, heat transport is enhanced dramatically and the local effective scaling exponent is close to 1/2, extending more than one decade. We call this Regime I, the enhanced exponent regime. This scaling exponent is very robust as it does not change when altering the roughness height in the range [0.05; 0.15]. The higher the roughness is, the earlier the system can step into Regime I. However, further increasing $\textrm{Ra}$ does not result in an extension of Regime I. Instead, the scaling exponent saturates back to the effective value $\beta \approx 0.33$, which is the typical Malkus exponent in the classical regime where the BL is of laminar type \cite{malkus1954,grossmann2000,grossmann2001}. We call this Regime II, the saturated exponent regime. Remarkably, the heat transport follows exactly the same line in this regime for different roughness height. The heat transfer increases 3.05 times while the wet surface area augment is 2.30 times, suggesting that the heat transfer enhancement is mainly due to the enlarged surface area while strong plume ejections in the cavities contribute the remaining part.

Next, we vary the roughness wavelength $\lambda$ as well, focusing on the effective scaling exponent $\beta$, up to $\mathrm{Ra}=10^{12}$. No matter what $\lambda$ is, we can still identify the Regime I where the effective exponent increases and Regime II where it saturates back to a value close to 0.33. Fig. \ref{fig3}(a) demonstrates the scaling exponents in Regime I. For each roughness height, there is always an optimal $\lambda$ which maximizes the effective scaling exponent to 1/2, which is the strict upper limit of scaling exponent in RB \cite{kraichnan1962,grossmann2011,goluskin2016}. However, for each $h$, the optimal $\lambda$ is different. A better parameter to describe the effects of roughness on the scaling exponent is the roughness aspect ratio $\lambda/h$, as shown in Fig. \ref{fig3}(b)(c) for Regime I and Regime II, respectively. Interestingly, all the data collapse into one line and specifically for the optimum we find $\lambda/h \approx 1$, irrespective of the roughness height. This insight offers the opportunity to vary the roughness features to {\it tune} effective scaling exponents in Regime I, in the large range [0.29; 0.50] and in Regime II, in the range [0.29; 0.33].        

\begin{figure}[htbp]

     \includegraphics[width=3.in]{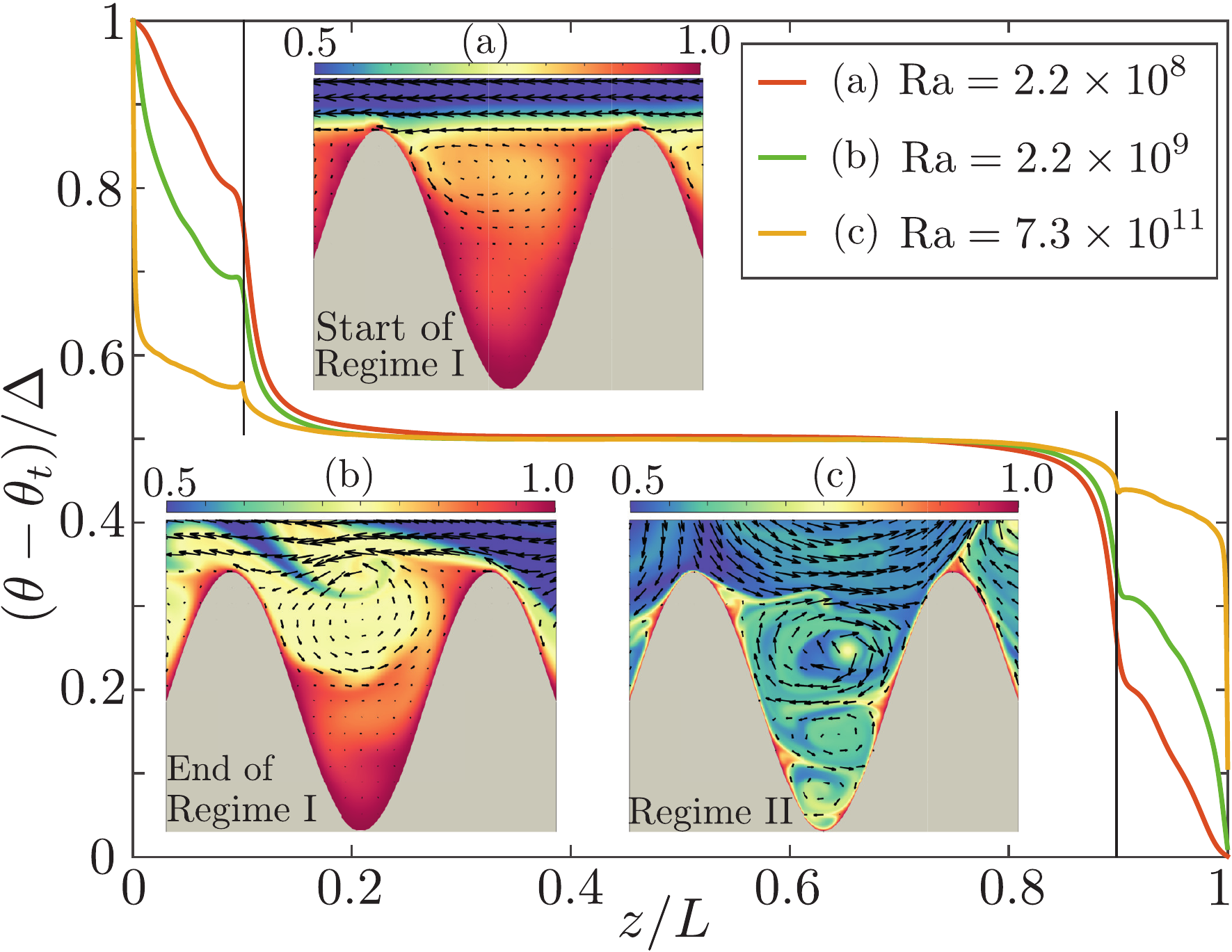}
    
      \caption{The dimensionless mean temperature profiles $(\theta-\theta_t)/\Delta$ for $\lambda/h=1.00$ and $h/L=0.10$ at (a) the start of Regime I ($\mathrm{Ra}=2.2 \times 10^8$), (b) the end of Regime I ($\mathrm{Ra}=2.2\times 10^9$) and (c) Regime II ($\mathrm{Ra}=7.8\times10^{11}$), where $\theta_t$ is the temperature of the top plate. The insets show the temperature fields, superposed by the velocity vectors in the cavity regions. The three insets share the same colormap. In Regime I, we observe one roll inside the cavity, whereas in Regime II, there are multiple rolls. The two black lines indicate where the tips of the roughness elements are.}
\label{fig4}
\end{figure}

The various studies which have been reported in the literature fall into either of the two regimes we revealed here. Namely, the regime where the effective scaling increases up to $\mathrm{Ra^{1/2}}$ \cite{roche2001,toppaladoddi2017} due to roughness or the regime where the scaling is similar \cite{shen1996,du2000,shishkina2011} to the smooth case. These seemingly contradictory viewpoints have caused some confusions in the interpretation of the data on RB convection with roughness. The present study has bridged the gap between the two views by studying a large enough regime in $\mathrm{Ra}$ and also various roughness characteristics. The clear conclusion is that the observed effective 1/2 scaling in Regime I should not be interpreted as the attainment of the so-called ultimate regime as suggested in previous studies \cite{roche2001,toppaladoddi2017}, but rather as a crossover regime in which the roughness elements start to perturb the thermal BL. This provides a consistent and plausible explanation for the observed scatter in the reported values of $\beta$ with the presence of roughness in prior studies \cite{ciliberto1999,qiu2005,stringano2006,tisserand2011,wei2014,salort2014,sebastian2014}, where different combinations of $h$ and $\lambda$ were chosen. We showed that tuning $h$ and $\lambda$ can lead to big variations of $\beta$, especially in Regime I (Fig. \ref{fig3}), presumably resulting in the scattered effective scaling exponents in previous experiments and numerical simulations.

To further disentangle the mechanisms leading to the two regimes, in Fig. \ref{fig4} we show the mean temperature profiles as well as the local flow structures inside the cavities for $\lambda/h=1.00$ and $h/L=0.10$ at different $\mathrm{Ra}$. We observe secondary vortices induced by large scale rolls. In Regime I, the weak secondary vortices can not efficiently mix the fluid in the cavities and thus the flow there is still viscosity dominated. Therefore, the temperature profile in the cavity is rather linear. In contrast, in Regime II, secondary vortices are strong enough to induce smaller vortices, which further induce even smaller vortices down to the valleys of the cavities, forming a cascade of vortices. Due to the strong mixing caused by this process, the roughness elements are covered by a thin thermal BL which is uniformly distributed along the rough surfaces, effectively mimicking an enlarged surface area. As a result, the mean velocity profile is steep only at the valley point of the cavity and otherwise becomes very similar to the smooth case.

\begin{figure}[htbp]

     \includegraphics[width=3.in]{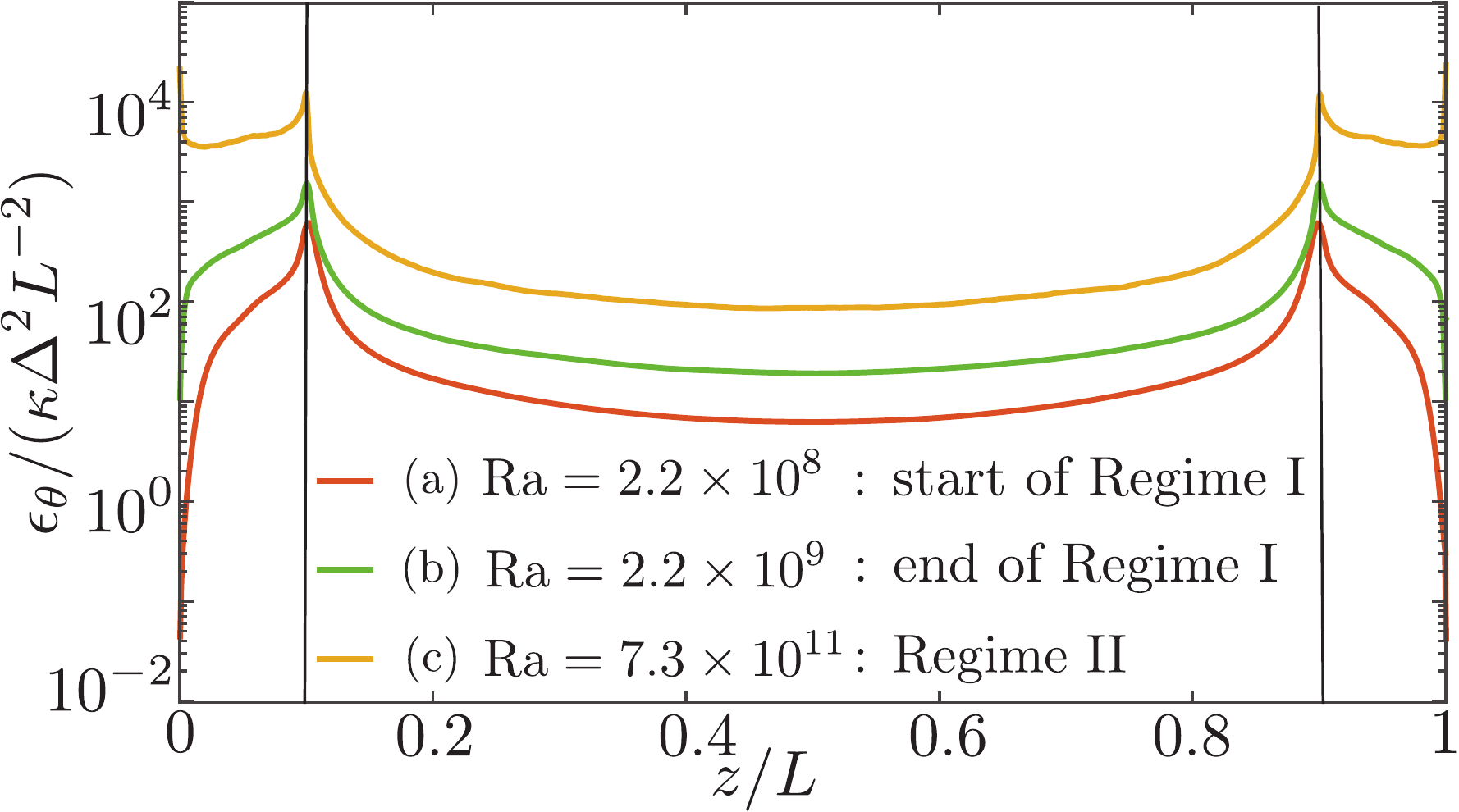}
    
      \caption{The dimensionless mean thermal energy dissipation rate $\epsilon_\theta/(\kappa\Delta^2L^{-2})$ across the height of the domain for $\lambda/h=1.00$, $h/L=0.10$ at (a) the start of Regime I ($\mathrm{Ra}=2.2 \times 10^8$), (b) the end of Regime I ($\mathrm{Ra}=2.2\times 10^9$) and (c) Regime II ($\mathrm{Ra}=7.8\times10^{11}$). Note that the thermal energy dissipation is shown on a log-scale. The two black lines indicate where the tips of the roughness elements are.}
\label{fig5}
\end{figure}

Inside the cavities, in Regime I, the viscosity dominated flow decreases the BL contribution to the total thermal energy dissipation rate, as compared to the smooth case, while in Regime II, the cascade of vortices and the restoration of the uniformly distributed thermal BL bring back the BL contribution to the total thermal energy dissipation rate. For the thermal energy dissipation rate, it has been well known that if the bulk contribution is dominant, the scaling exponent is close to $1/2$ and if the BL contribution dominates, the scaling exponent is close to $1/3$, i.e., in the classical regime where the BL is of laminar type \cite{grossmann2000,grossmann2001}. Here, due to the effective scaling, Regime I seems to be the bulk dominated regime whereas Regime II seems to be the classical BL controlled regime. This is counterintuitive since one would expect the opposite with increasing $\mathrm{Ra}$ for the smooth RB, i.e. the system can only become more bulk dominated with increasing $\mathrm{Ra}$ \cite{grossmann2000,grossmann2001,grossmann2011}. In Fig. \ref{fig5} we show the mean thermal energy dissipation rate along the height between the two plates. Indeed, in Regime I, the thermal dissipation inside the cavity is negligible and thus it is bulk dominated, whereas in Regime II, the thermal dissipation inside the cavity is dominated and thus it corresponds to the classical BL controlled regime, supporting above interpretation on the reverse role of BL and bulk in the presence of roughness.

In conclusion, the present study has demonstrated that the local effective $\beta=1/2$ scaling in RB with roughness does not necessarily indicate the start of the ultimate regime as claimed in previous studies \cite{roche2001,toppaladoddi2017}. Instead, its observation is fortuitous because by tuning the height and wavelength of roughness elements simultaneously, $\beta$ can be tuned between 0.29 and 0.5 locally. This Regime I is just a crossover regime where the bulk is dominated, as has been speculated in Ref. \cite{ahlers2009,tisserand2011,xie2017}. Further increasing $\mathrm{Ra}$ brings back the thin BL inside the cavities and restores the classical BL controlled regime, which also causes the scaling to saturate and recover the classical RB scaling exponent. Only once the BLs become turbulent does the transition to the ultimate regime occur \cite{ahlers2009, grossmann2011}.   

Finally, we note that for Taylor-Couette flow with grooved roughness which align with the azimuthal flow direction, our previous DNS showed that for the dimensionless torque scaling $\mathrm{Nu_\omega} \sim \mathrm{Ta}^\beta$, both Regime I where $\beta$ increases up to 1/2 and Regime II where $\beta$ saturates back were also observed \cite{zhu2016a}. Here, $\mathrm{Ta}$ is the dimensionless angular velocity difference which plays the equivalent role to $\mathrm{Ra}$ in RB. Thus, there is strong evidence that the two systems are not only analogous with each other in the smooth case \cite{eckhardt2007a,eckhardt2007b,vangils2011,he2012b,grossmann2016} but also in the rough case, corroborating the conclusions of the current study from a broader point of view.

%





\begin{acknowledgments}
We thank V. Mathai for fruitful discussions. This work is supported by FOM and MCEC, both funded by NWO. We thank the Dutch Supercomputing Consortium
SURFSara, the Italian supercomputer Marconi-CINECA through the PRACE Project
No. 2016143351 and the ARCHER UK National Supercomputing Service through the
DECI Project 13DECI0246 for the allocation of computing time.
\end{acknowledgments}


\bibliography{RB_rough}

\end{document}